# Modeling Vascular Branching Alterations in Polycystic Kidney Disease

Timothy L. Kline[1,2]

[1]*Department of Radiology and* [2]*Division of Nephrology and Hypertension,*
*Mayo Clinic, 200 1st Street SW, 55905, USA*

The analysis of biological networks encompasses a wide variety of fields from genomic research of protein-protein interaction networks, to the physiological study of biologically optimized tree-like vascular networks. It is certain that different biological networks have different optimization criteria and we are interested in those networks optimized for fluid transport within the circulatory system. Many theories currently exist. For instance, distributive vascular geometry data is typically consistent with a theoretical model that requires simultaneous minimization of both the power loss of laminar flow and a cost function proportional to the total volume of material needed to maintain the system (Murray's law). However, how this optimized system breaks down (or is altered) due to disease has yet to be characterized in detail in terms of branching geometry and geometric interrelationships. This is important for understanding how vasculature remodels under changes of functional demands. For instance, in polycystic kidney disease (PKD), drastic cyst development may lead to a significant alteration of the vascular geometry (or vascular changes may be a preceding event). Understanding these changes could lead to a better understanding of early disease as well as development and characterization of treatment interventions. We have developed an optimal transport network model which simulates distributive vascular systems in health as well as disease in order to better understand changes that may occur due to PKD. We found that reduced perfusion territories, dilated distributive vasculature, and vessel rarefaction are all consequences of cyst development derived from this theoretical model and are a direct result of the increased heterogeneity of local renal tissue perfusion demands.

Graph theory, the study of network structures in terms of vertices and edges, is an applicable approach in disparate research studies; from the study of protein-protein interaction networks [1, 2], to the study of vascular networks [3, 4]. For instance, the branching geometry of vascular trees is investigated by analyzing the geometric properties of different vascular structures [5, 6]. In this sense, the vascular branching geometry is modeled as a collection of graph edges (interbranch segments) and nodes (bifurcation points), and properties such as interbranch segment length and diameter are measured. The experimental data can then be compared to theoretical models of optimal transport networks [7, 8] such as that derived by Murray [9]. Murray's law is based on the concept that the geometry of the fluid transport system (i.e. vasculature) is consistent with simultaneous minimization of both the power loss of laminar flow and of a cost function proportional to the total volume of material needed to maintain the system (i.e. lumenal contents) - factors that have opposing geometric consequences. Knowledge from vasculature has been used in the design of optimized microfluidic channels [10] and synthetic vessel constructs [11].

Utilizing an optimal transport network modeling approach (grounded in graph theory), random fluctuations have been attributed to the loop-like venation in leaves [12, 13]. Here, individual nodes are modeled to behave as current sources (an analogy to electrical transport networks), the leaf's root is modeled as a current drain (i.e., the source of nutrients or pathway for waste removal), and the connections (edges) between the nodes carry a variable current. The conductance (which is related to the diameter of an edge) is modeled to minimize the network's dissipation, while constrained to do so utilizing a limited amount of material. The occurrence of loop-like networks in the case of varying current sources suggests that random attacks (such as bugs feeding on the leaf) are partially overcome by the optimal transport network being composed of multiple pathways to a single perfusion region. Thus, in certain biological systems, it is suggested that loops may form which thereby serve to increase the robustness of the transport network.

Here we develop an optimal transport network model in order to study distributing vascular systems subjected to varying degrees of heterogeneous demands in order to simulate cystic development in polycystic kidney disease (PKD). We hypothesize that this model will help further our understanding of disease-based vascular changes which may precede other structural and functional changes and thereby serve as early disease biomarkers.

We consider a formulation of the optimal transport network problem which involves the minimization of the total dissipation rate, under the constraint that a limited amount of material is available [14]. In graph theory, a network can be modeled as a set of vertices or nodes $k$, and a set of edges or connections between the nodes, $(k, l)$. Our analogous system is an electrical transport network where each node acts as either a current source or drain (exact arrangement discussed later), and a variable current $I_{kl}$ flows through the edges. Each edge has an associated conductance $\kappa_{kl}$, and a length $L_{kl}$. The dissipation rate $J$ is related to the currents and conductances by

$$J = \sum_{(k,l)} \frac{I_{kl}^2}{L_{kl}\kappa_{kl}}. \qquad (1)$$

The dissipation rate is then minimized through the currents and conductances with the local constraint that $i_k = \sum_l I_{kl}$ (Kirchhoff's current law), and a global constraint that the

sum of the conductances raised to a given power is kept constant, as in

$$K^\gamma = \sum_{(k,l)} \kappa_{kl}^\gamma. \quad (2)$$

$\kappa_{kl}^\gamma$ can be recognized as the building cost of a channel (i.e,. a vessel branch segment) and the variable $\gamma$ depends on the nature of the network. As discussed by Corson [12], for the case of electrical wires, $\gamma = 1$, and for a model based on Murray's law, $\gamma = 0.5$. In the case of Murray's law, the total luminal contents (i.e. surface area of blood vessels) is assumed fixed, thus, $\sum_{(k,l)} V = K^{0.5}$. Also, $\gamma > 1$ is of little relevance since in this case it is more economical to build several parallel links having a small conductance rather than a large one of equivalent capacity. Note that this is necessary in the case that distinct elements need to be carried between certain nodes (e.g., what may occur if planning a city street layout). For $\gamma < 0$ the model is degenerate, and thus $\gamma$ is said to lie between 0 and 1, though network properties regarding the network's phase transition when $\gamma$ crosses 1 have been explored [14].

Using a Lagrange multiplier $\lambda$, we define the minimization problem as

$$\Xi(\kappa kl, Ikl) = \sum_{(k,l)} \frac{I_{kl}^2}{L_{kl}\kappa_{kl}} - \lambda \sum_{(k,l)} \kappa_{kl}^\gamma. \quad (3)$$

The minima of the dissipation, with constant K has the following necessary conditions

$$\frac{\partial \Xi}{\partial I_{kl}} = 0, \frac{\partial \Xi}{\partial \kappa_{kl}} = 0. \quad (4)$$

Solving for the derivative of $\Xi$ with respect to $\kappa_{kl}$, Bohn et al [14] gave an explicit scaling relation between the currents and the conductivity in the minimal configuration

$$\kappa_{kl} = \frac{K(I_{kl}^2/L_{kl})^{1/(1+\gamma)}}{\left(\sum_{(m,n)}(I_{kl}^2/L_{kl})^{1/(1+\gamma)}\right)^{1/\gamma}}. \quad (5)$$

This is the equation used to solve for each edge capacitance (i.e. derive the optimal branching geometry in terms of vessel diameters and arrangement).

To set up our network model, we begin with a collection of $n^2$ nodes, arranged in an $n \times n$ grid. Our problem is formulated in $\mathbb{R}^2$ meaning that each node is connected to at most 4 adjacent nodes (side nodes are connected to 3 other nodes, and corner nodes are connected to 2 other nodes) through edges which have the conductance $\kappa_{kl}$ and carry the current $I_{kl}$ between node $k$ and node $l$. The length of each edge $L_{kl}$ is here set to unity. The network consists of sources $s$ and drains $d$, where $s + d = n^2$ and $\sum i_s = -\sum i_d$. The initial condition consists of a random distribution of conductances. Then, the potential $U$ at the individual nodes are found by solving the system of linear equations $i_k = \sum_l \kappa_{kl}(U_k - U_l)$ (which are solved by noting that $AX = B$ in matrix notation). The currents $I_{kl}$ can then be determined and plugged in to Eq. 5 to determine the new conductances. These new conductances are then used in the next iteration of the minimization algorithm. These steps are repeated until the values of the conductivity have converged.

In order to understand how the changes in vascular branching likely impact regional perfusion, we modeled the flow properties of the vascular network. A map was created that contains information pertaining to the resistance to flow at each pixel within the network. The map assigns a value to each pixel by use of the Hagen-Poiseuille equation [15]

$$\Delta P = \frac{8\mu \cdot l \cdot Q}{\pi \cdot r^4}, \quad (6)$$

which describes slow viscous incompressible flow along a tubular pathway. $\Delta P$ is the pressure difference between an inlet and outlet, $\mu$ is the dynamic viscosity of the fluid, $l$ is the length of the path, $Q$ is the volumetric flow rate, and $r$ is the radius along the path. The map's value at each pixel therefore characterizes the resistance by

$$R_{mn} = \sum \frac{l}{r^4}, \quad (7)$$

where $l$ is the length, $r$ is the radius, and $R_{mn}$ is the determined resistance value at pixel (m,n).

To create the map we have implemented the fast marching method [16], using a starting point at the vessel tree root. The fast marching method is a numerical scheme for solving the Eikonal equation

$$|\nabla u_{mn}| = F_{mn}, \quad (8)$$

where $u$ is the arrival time function and $F$ corresponds to the weight function or the speed of the front progression. The method relies on an approximation to this gradient given by

$$\begin{bmatrix} \max(D_{mn}^{-x}u, -D_{mn}^{x}, 0)^2 + \\ \max(D_{mn}^{-y}u, -D_{mn}^{y}, 0)^2 \end{bmatrix}^{1/2} = F_{mn}, \quad (9)$$

where the $D_{mn}$ are represented by,

$$D_{mn}^{-x}u = \frac{u_j - u_{j-1}}{h}, D_{mn}^{+x}u = \frac{u_{j+1} - u_j}{h}. \quad (10)$$

Here h is the pixel's length along $m$ (relations are similar for $D$ in the $n$ direction). By using a weighting function going as $r^{-4}$, the difficulty of perfusing a particular region, based on network geometry, can be obtained.

Shown in Fig. 1 are the results for the computed optimal transport network composed of 30 x 30 nodes (panel A) and interconnections (panel B) used to simulate a distributive vasculature system supplying rather homogeneous tissue (locations with nutrient supply/waste disposal demands held constant during optimization). In all cases, $\gamma$ was set to 0.5 (to simulate vascular networks characterized by Murray's law and bounded to have a constant vessel surface area). The individual sources were initialized to all have randomly assigned current values, and the single drain located at the bottom left node, was initialized to have the negative of the sum of all other source nodes. Two example vessel trees are shown (panel C and panel D), highlighting how the random initialization results in very different network architectures.

Shown in Fig. 2 are the results for the computed optimal transport network again composed of 30 x 30 nodes used to simulate a distributive vasculature system supplying heterogeneous tissue demands due to cystic development. Vascular network with no cysts is shown (panel A), as well as with cysts with increased demand (panel B).

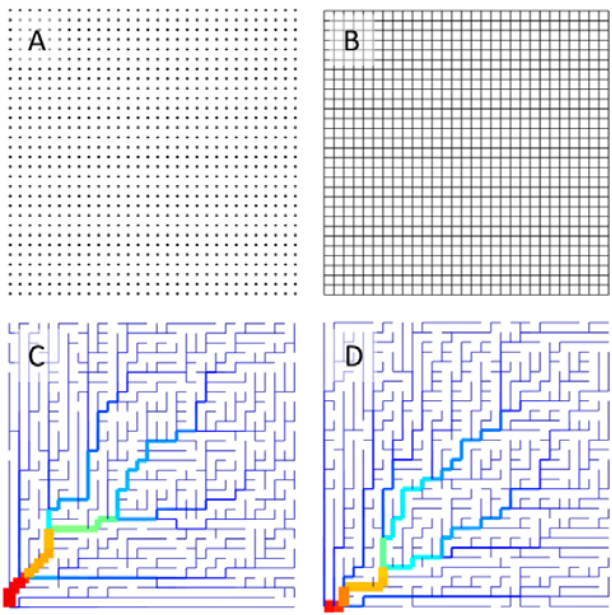

FIG. 1. Network initialization and example vascular networks. Panel A: network nodes used to simulate local demands for supply of nutrients/waste disposal. Panel B: network edges used to simulate the available vascular network to be optimized by our model system. Panels C and D: example vascular networks generated by the algorithmic approach. Due to random initialization of the source nodes, rather different network architectures can result. Colormap value and thickness of the segments correspond to branch diameter.

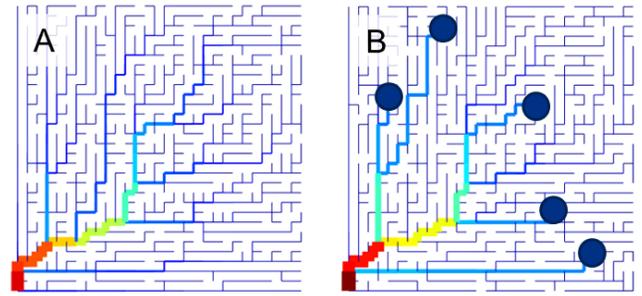

FIG. 2. Panel A: vascular network with no cysts. Panel B: vascular network with 5 cystic regions with a 15x local demand on supply. Notice how vasculature dilates in response to the increased need to supply certain regions more than others.

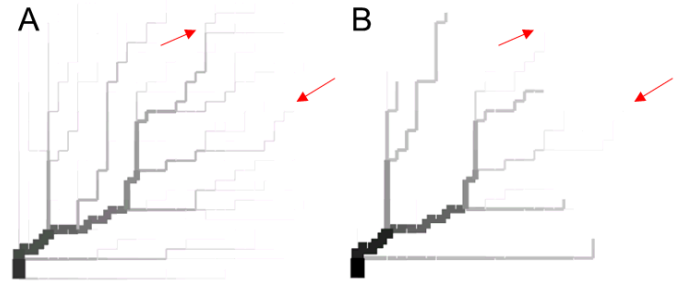

FIG. 3. Vessel rarefaction is a consequence of cystic burden and is elucidated by our model. Panel A: no cysts. Panel B: 5 cysts with 15x demand. Red arrows highlight vascular regions lost due to the more heterogeneous tissue demand of the cystic regions. The darker and thicker the branch segment, the larger the diameter.

In addition, vessel rarefaction is seen to increase as a result of an increased demand from cystic development. Shown in Fig. 3 are network architectures for no cysts, as well as 5 cysts requiring 15x tissue demand.

As demand from cysts increases, the overall efficiency of the network in non-cystic regions (locations away from cysts) is seen to decrease. Shown in Fig. 4 are reduced perfusion territories resulting from cystic development.

Autosomal dominant polycystic kidney disease (ADPKD) is the most common genetic disorder involving a single gene and is the fourth leading cause of end-stage renal disease (ESRD) [17, 18]. Despite the renal complications associated with cyst formation compromising renal function, cardiovascular disease is the main cause of morbidity and mortality in ADPKD patients [19]. In addition, other extra-renal manifestations such as intracranial aneurysms, subarachnoid hemorrhage, and spontaneous cervicocephalic artery dissections may cause debilitating injury and often premature death [20-23]. The importance of our understanding of the vascular phenotype of ADPKD is thus crucially important.

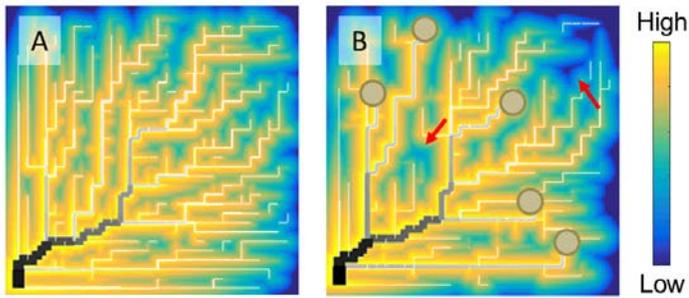

FIG. 4. Reduced perfusion territories are observed in the networks with cystic burden. Panel A: no cysts. Panel B: 5 cysts with 15x demand. Red arrows highlight regions of reduced perfusion resulting from vessel rarefaction/shrinking.

Currently, endothelial dysfunction is the earliest observable manifestation of ADPKD [24, 25]. Oxidative stress and vascular inflammation have been linked to this vascular dysfunction which includes increased contraction and decreased relaxation of the renal distributive vasculature, which results in tissue ischemia (a stimulus for angiogenesis of the exchange vasculature of the kidneys) [21]. Evidence of angiogenesis on the surface of renal cysts has been shown and high levels of angiogenic growth factors including vascular endothelial growth factor (VEGF) have been reported in cyst fluid and in the circulation system [26]. Other notable findings include the expression of polycystin (the large protein encoded by the PKD1 and PKD2 genes) in arterial smooth muscle [27], defective nitric oxide generation from diminished vasodilator cNOS activity [28, 29], up-regulation of the endothelin isoform ET-1 contributing to vasoconstriction [30], and high levels of lipoprotein(a) which more than likely contributes to the high incidence of cardiovascular events in ADPKD [24].

Parameters known to antedate the decrease in renal function of ADPKD patients include renal structure, renal blood flow (RBF), and mean arterial pressure (MAP) [31, 32]. Renal blood flow reduction has been shown to parallel total kidney volume (TKV) increases, to precede decreases in glomerular filtration rate (GFR), and to predict structural and functional disease progression [33]. The proliferation of renal epithelial cells and the formation and growth of cysts that replace normal parenchyma of the kidney suggests that a great deal of remodeling and expansion of the vasculature must occur to provide oxygenation and nutrition to the cyst cells. Well-defined vascular networks surrounding cysts, dilated capillaries, as well as the loss of normal vascular architecture have been shown in scanning electron microscopy studies of vascular casts [34]. In addition, decreased vascular densities have been revealed by micro-CT where the vasculature of kidney samples was injected with a lead-based polymer [35]. Evidently, a debilitating feedback loop promoting allocation of vascular maintenance away from renal vascular supporting healthy tissue is indicated.

In this current study, we modeled the expected changes that should occur in response to the demands of cystic development. Dilation of the distributive vasculature results from the optimal transport modeling approach used in this study, as well as vessel rarefaction and reduced perfusion territories. We believe that providing models to characterize vascular changes occurring due to disease will facilitate a better understanding of disease mechanisms and help in developing earlier disease biomarkers.

This study was supported in part by the Mayo Clinic Robert M. and Billie Kelley Pirnie Translational PKD Center and the NIDDK grants P30DK090728 and K01DK110136.